\documentclass{JINST}

\def\ifmath#1{\relax\ifmmode #1\else $#1$\fi}%
\def\pbi{\mbox{pb$^{-1}$}}%
\def\fbi{\mbox{fb$^{-1}$}}%

\def\pt{\ifmath{\mathrm{p_T}}}

\title{Uniformity and Stability of the CMS RPC Detector at the LHC}

\author{S. Costantini$^a$\thanks{Corresponding author.},
K. Beernaert$^a$, A. Cimmino$^a$, G. Garcia$^a$, 
J. Lellouch$^a$,
A. Marinov$^a$, A. Ocampo$^a$, N. Strobbe$^a$, F. Thyssen$^a$, 
M. Tytgat$^a$, P. Verwilligen$^a$\thanks{Now at Universit\`a e INFN, 
Sezione di Bari.},
E. Yazgan$^a$, N. Zaganidis$^a$, 
A. Dimitrov$^b$, 
R. Hadjiiska$^b$, L. Litov$^b$, B. Pavlov$^b$, P. Petkov$^b$, 
A. Aleksandrov$^c$, V. Genchev$^c$, P. Iaydjiev$^c$, M. Rodozov$^c$, 
M. Shopova$^c$, G. Sultanov$^c$, 
Y. Ban{$^d$}, J. Cai{$^d$}, 
Y. Ge{$^d$}, Q. Li{$^d$}, 
S. Qian{$^d$}, Z. Xue{$^d$}, 
C. Avila{$^e$}, L.F. Chaparro{$^e$}, J.P. Gomez{$^e$}, B. Gomez Moreno{$^e$}, 
A.F. Osorio Oliveros{$^e$}, J.C. Sanabria{$^e$}, 
Y. Assran{$^f$}, 
A. Sharma{$^g$}, 
M. Abbrescia{$^h$}, C. Calabria{$^h$}, A. Colaleo{$^h$}, 
F. Loddo{$^h$}, M. Maggi{$^h$}, G. Pugliese{$^h$}, 
L. Benussi{$^i$}, S. Bianco{$^i$}, S. Colafranceschi{$^i$}, D. Piccolo{$^i$}, 
S. Buontempo{$^j$}, C. Carrillo{$^j$}, O. Iorio{$^j$}, P. Paolucci{$^j$}, 
U. Berzano{$^k$}, M. Gabusi{$^k$}, P. Vitulo{$^k$}, 
M. Kang{$^l$}, K.S. Lee{$^l$}, S.K. Park{$^l$}, S. Shin{$^l$},
Y. Choi{$^m$}, J. Goh{$^m$}, M.S. Kim{$^m$},  and H. Seo{$^m$} \\
\\
\llap{$^a$}University of Ghent, Department of Physics and Astronomy, \\
	Proeftuinstraat 86, B-9000, Ghent, Belgium\\
\llap{$^b$}University of Sofia, Faculty of Physics, Atomic Physics 
Department,\\
  5, James Bourchier Boulevard, BG-1164 Sofia, Bulgaria\\
\llap{$^c$}Bulgarian Academy of Sciences,\\
  Inst. for Nucl. Res. and Nucl. Energy,\\
  Tzarigradsko shaussee Boulevard 72, BG-1784 Sofia, Bulgaria\\
\llap{$^d$}Peking University, School of Physics,\\
  CN-100871 Beijing, China\\
\llap{$^e$}Universidad de Los Andes,\\
  Apartado A\'ereo 4976, Carrera 1E, no. 18A 10, CO-Bogot\'a, Colombia\\
\llap{$^f$}Academy of Scientific Research and Technology of the Arab 
Republic of Egypt,\\
  101 Sharia Kasr El-Ain, Cairo, Egypt\\
\llap{$^g$}Panjab University, Department of Physics,\\
  Chandigarh Mandir 160 014, India\\
\llap{$^h$}Universit\`a e INFN, Sezione di Bari,\\
  Via Orabona 4, IT-70126 Bari, Italy\\
\llap{$^i$}INFN, Laboratori Nazionali di Frascati (LNF),\\
  PO Box 13, Via Enrico Fermi 40, IT-00044 Frascati, Italy\\
\llap{$^j$}Universit\`a e INFN, Sezione di Napoli,\\
  Complesso Univ. Monte S. Angelo, Via Cintia, IT-80126 Napoli, Italy\\
\llap{$^k$}Universit\`a e INFN, Sezione di Pavia,\\
  Via Bassi 6, IT-Pavia, Italy\\
\llap{$^l$}Korea University, Physics Department,\\
  Seoul Cheongryangri 143-701, Republic of Korea\\
\llap{$^m$}Sungkyunkwan University, Department of Physics,\\
  2066, Seobu-ro, Jangan-gu, Suwon, Gyeonggi-do, Republic of Korea\\

  E-mail: \email{Silvia.Costantini@cern.ch} \\
}

\abstract{The Resistive Plate Chambers (RPCs) are employed in the 
CMS experiment
at the LHC as dedicated trigger system both in the barrel and in the endcap.
This note presents results of the RPC detector uniformity and stability during the 
2011 data taking period, and preliminary results obtained with 2012 data.
The detector uniformity has been ensured with a dedicated High Voltage
scan with LHC collisions, 
in order
to determine the optimal operating working voltage of each individual
RPC chamber installed in CMS.
Emphasis is given on the procedures and results of the High Voltage calibration.
Moreover, an increased detector stability has been obtained by automatically taking into
account temperature and atmospheric pressure variations in the CMS cavern. 
}


\begin{document}

\section{The CMS experiment at the LHC}
\label{section:intro}

The Large Hadron Collider (LHC) has become operational in 2009. High-energy 
physics runs took place in 2010 and 2011,
with proton-proton 
collisions at a center-of-mass energy of 7 TeV,
and in 2012,
with proton-proton 
collisions at a center-of-mass energy of 8 TeV.
The maximal instantaneous luminosity reached
$3.5 \cdot 10^{33}$ cm$^{-2}$ s$^{-1}$ in 2011 and almost twice
this value in 2012 at the time this article was written.

The Compact Muon Solenoid (CMS)~\cite{cms1, cms2} Collaboration, 
one of the six experiments 
currently operating at the LHC, consists of over 3000 scientists, engineers and
graduate students from 173 institutes in 40 Countries.


The central feature of the 
CMS detector 
is a superconducting solenoid,
of 6 m internal diameter, providing a field of 3.8 T. 
Within the field volume are the silicon pixel and strip tracker, 
the lead-tungstate crystal electromagnetic calorimeter, and the
brass-scintillator hadron calorimeter. Muons are measured in 
gas-ionization detectors embedded in the steel return yoke. 
In addition to the barrel and endcap detectors, CMS has
extensive forward calorimetry, assuring very good hermeticity
with pseudorapidity coverage up to high values ($|\eta| < 5 $).


Muons with pseudorapidity in the range $| \eta | < 2.4 $ are measured 
with detection planes made of three technologies: 
Drift Tube chambers (DT), Cathode Strip Chambers (CSC)
and Resistive Plate Chambers (RPC). 
Matching the muons 
to the tracks measured in the silicon tracker gives 
a transverse momentum (\pt) resolution between 1\% and 5\%, for $\pt$ values up 
to 1 TeV.


In 2011, CMS has recorded 5.20~\fbi of data out
of 5.72~\fbi delivered by the LHC, for an efficiency of 91\%. Roughly 93\%
of the recorded data has been certified as ``golden'' for all physics analyses.
An average of 98\% of the subdetector channels are operational and in the
readout. 
In 2012, at the time this article was submitted, 9.98~\fbi of data out
of 10.66~\fbi delivered had already been recorded.

\section{The CMS RPCs}

\begin{figure}[hbp]
\includegraphics[width=.49\textwidth]{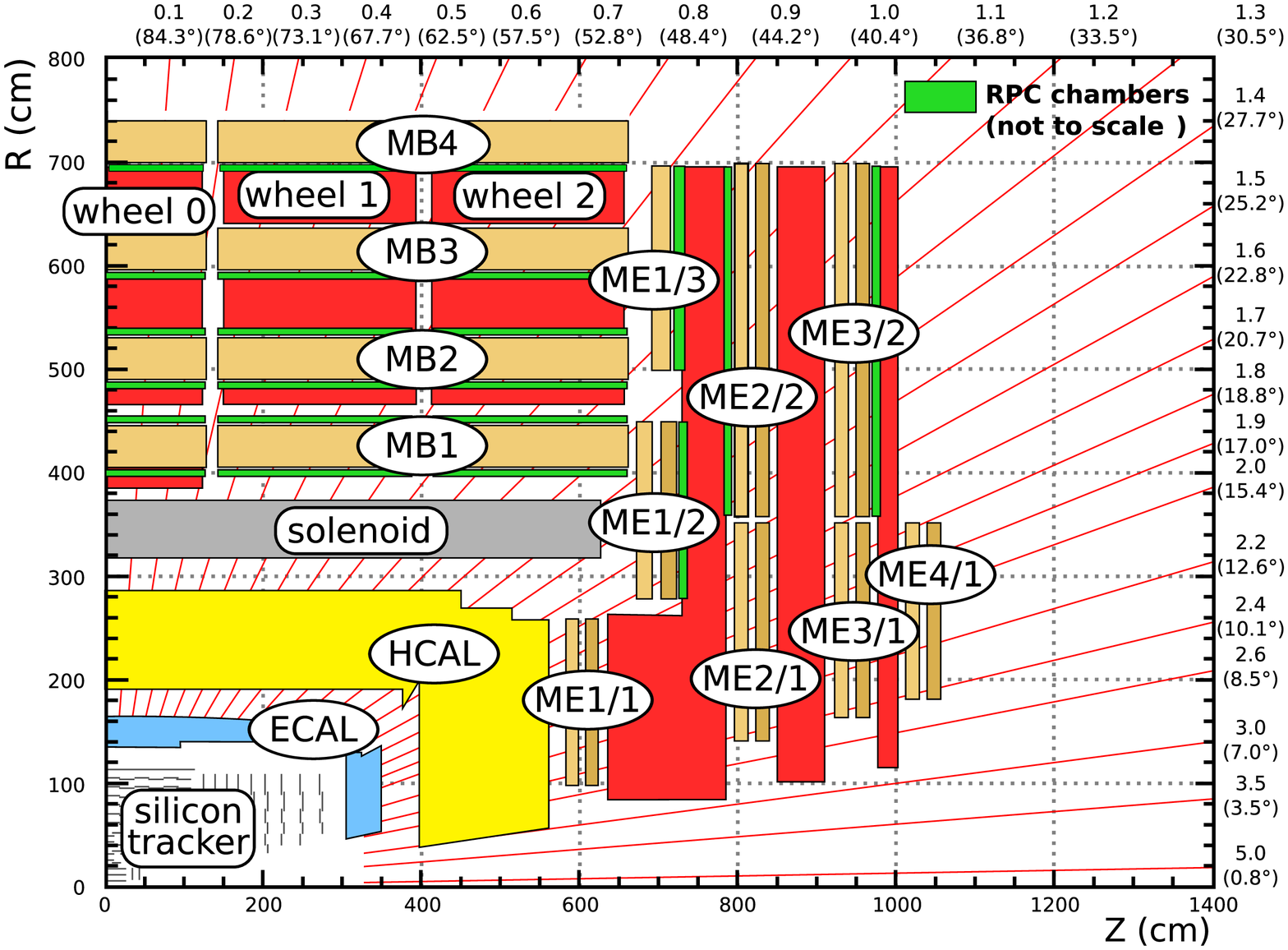}
\includegraphics[width=.49\textwidth]{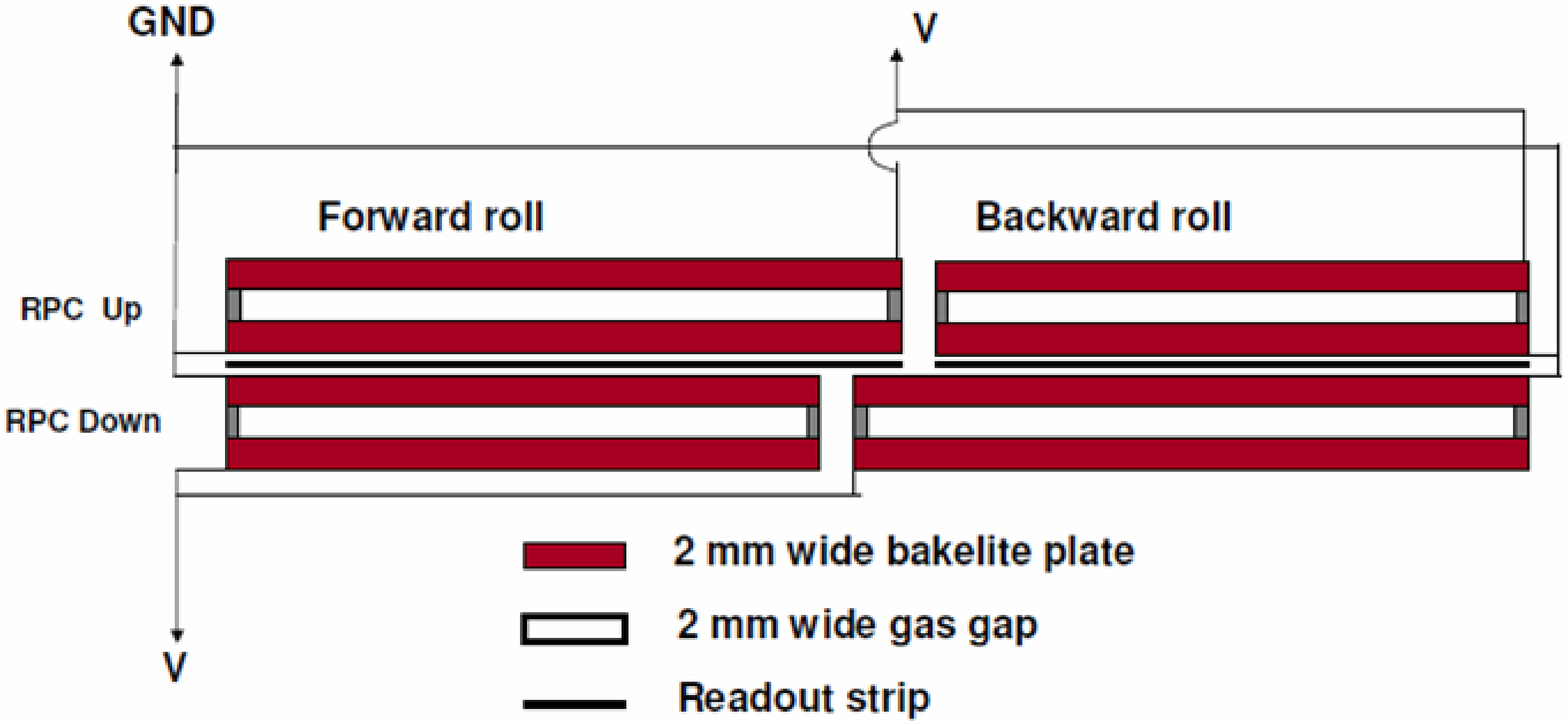}
\caption{Left: Schematic view, in the R-z plane, of one quadrant of the CMS detector.
The position of the RPC chambers is shown (not to scale).
Right: Schematic layout of a double-gap barrel chamber composed by two sub-units, called rolls.  
The readout strip plane is also shown. 
}
\label{fig:detector}
\end{figure}

The RPC~\cite{cms_rpc1, cms_rpc2} detectors are implemented in CMS 
as a dedicated trigger system, 
both in the barrel and in the endcap regions. 
Figure~\ref{fig:detector} shows a schematic view of 
one quarter of the CMS detector
in the R-z plane (Fig.~\ref{fig:detector}, left)
and the layout of a double-gap barrel chamber (Fig.~\ref{fig:detector}, right).
Two gas gaps, of 2 mm each, are formed by two parallel bakelite electrodes, with
one single plane of copper read-out strips  in-between. The two gaps feature
2 mm thickness and have a bulk resistivity of the order of $10^{10} \Omega$~cm.

The barrel RPC system consists of five wheels,
installed at $| \eta |< $ 0.8 and $| \mathrm{z} |< 7 $~m,
subdivided into 12 azimuthal sectors, each one equipped 
with six radial layers of RPCs. Six endcap disks, three on the positive and
three on the negative endcap side, are divided into 36 azimuthal sectors,
with two radial rings in each one. They assure a full coverage 
up to $| \eta |< $ 1.6. 
In total, 480 barrel chambers and
432 endcap chambers are installed, adding up to 68136 barrel strips and
41472 endcap strips, respectively, covering a total surface of about 3000 m$^2$. 
The CMS RPCs work in saturated avalanche mode and use a three-component,
non-flammable gas mixture composed of 95.2\% C$_2$H$_2$F$_4$ (R134a), 
4.5\% iC$_4$H$_{10}$ (isobutane) and 0.3\% SF$_6$.
Water vapor is added in order to maintain
the relative humidity at constant values and allow for 
constant bakelite resistivity.

\section{Resolution and efficiency studies: the method}
\label{section:method}

\begin{figure}[htbp]
\begin{center}
\includegraphics[width=.45\textwidth]{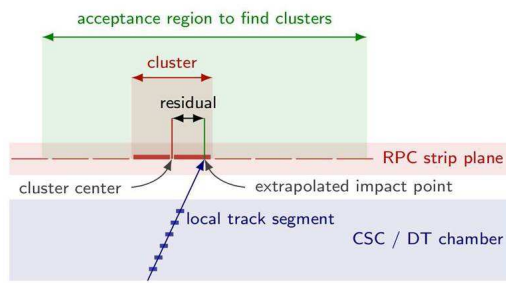} 
\caption{Schematic drawing of the extrapolation method. A DT or CSC
track segment is linearly extrapolated to the RPC strip plane. RPC hits
are sought in the acceptance region around the extrapolated impact point.
}
\label{fig:extrapolation}
\end{center}
\end{figure}

The presence, next to each RPC chamber,
of either a DT (in the barrel) or a CSC chamber (in the endcap), allows
to profit by the redundancy of the CMS muon system in order
to define the RPC hit efficiency independently of final physics ``objects''.
Muon track segments, reconstructed in the multi-layer detectors DT and CSC,
are linearly extrapolated to the RPC strip plane
and used to predict RPC hits in a fiducial region around the extrapolated
impact points, as illustrated in Figure~\ref{fig:extrapolation}.
Each extrapolated hit is matched to the closest RPC cluster, which
is formed by contiguous fired strips. 
The typical cluster size corresponds to 
about two strips, as shown in Section~\ref{section:stability}.

An additional requirement is applied to ensure that only extrapolated segments 
associated to muon tracks with hits in the central tracker
are taken into account, while DT/CSC segments 
probably originated by neutral background particles are discarded.
The three muon detectors employ different technologies and materials, thus
they have different sensitivities to the various backgrounds.

The extrapolation method allows to measure both the RPC spatial resolution, through
the residuals, and the hit efficiency.
The measured spatial resolution increases with increasing strip widths
and it ranges from 0.8 cm (for inner detector layers, characterized by smaller
strip pitches) to 1.3 cm (for outer
layers), both in the Barrel and in the Endcap. 
The strip pitch dimensions are between 2.28 and 4.10 cm in the barrel and between 
1.95 and 3.63 cm in the endcap.

\section{Detector uniformity. The High Voltage Scan} 
\label{section:scan}

High Voltage (HV) scans were performed at the beginning of the 2011 and of the 2012
proton-proton LHC running,
aiming at determining the optimal operating HV for each individual chamber.
 The variation of
the environmental pressure $\mathrm P$ and the temperature $\mathrm T$ inside the CMS cavern
was taken into account as described in Section~\ref{section:stability}.

Collision data were taken at eleven different values of the effective HV ($\mathrm{HV_{eff}}$,
defined in Eq.~(\ref{eqn:corr}) below), 
from 8.5~kV to 9.7~kV. It is worth mentioning that  only a 
negligible amount of data (about 3~\pbi \,  out of 5.72 \fbi in 2011, 
and about 6~\pbi in 2012) 
was discarded, {\it i.e.} not included in the ``golden'' sample 
for physics analyses, because of the RPC HV calibration.
%
A dedicated data stream was used, containing 
information from the muon detectors and the first level trigger.

The efficiency ($\epsilon$) dependence on $\mathrm{HV_{eff}}$ is parametrized 
by a sigmoidal
response function that can be written as:

\begin{equation}
\nonumber
\mathrm{
\epsilon = \frac{\epsilon_{max}} {1 + e^{-\lambda (HV_{eff} - HV_{50\%} )}}
}
\end{equation}
\noindent
where $\epsilon_{max}$ is the asymptotic efficiency for HV $\rightarrow 
\infty$, the 
$\lambda$ coefficient is proportional to the sigmoid slope at the inflection point, and
the High Voltage value HV$_{50\%}$ is the inflection point of the 
sigmoid, for which 50\% of  $\epsilon_{max}$ is reached.

\begin{figure}[htbp]
\begin{center}
\includegraphics[width=.75\textwidth]{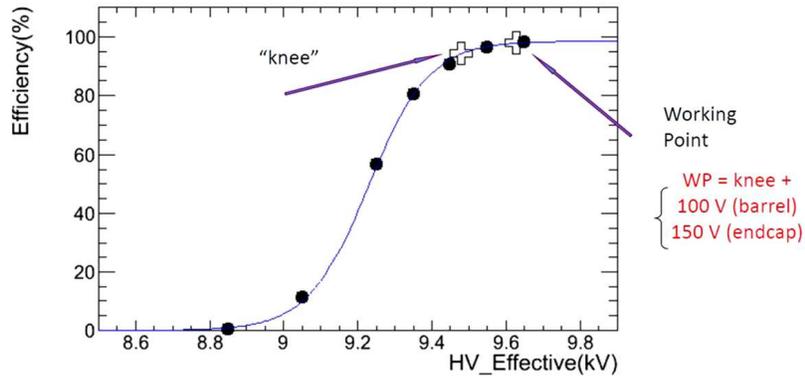}
\caption{
Efficiency measurement (in \%), as a function of the effective
High Voltage 
$\mathrm{HV_{eff}}$ (in kV) for a typical chamber. The ``knee'' and the
optimal working point, shown in the picture, are defined in the text. 
}
\label{fig:knee}
\end{center}
\end{figure}

\begin{figure}[htbp]
\begin{center}
\includegraphics[width=.65\textwidth]{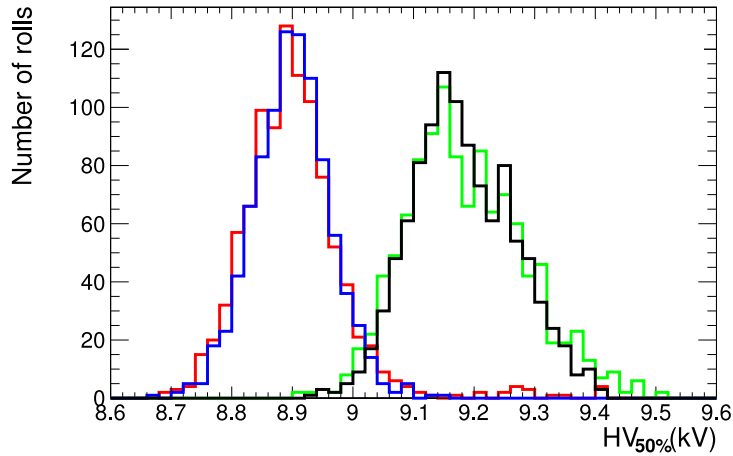} 
\caption{
HV$_{50\%}$ (in kV) distribution for the barrel (left curves, red and blue)
and the endcap (right curves, green and black) rolls. The red and green curves
show 2012 data.
HV$_{50\%}$ is the inflection point
of the sigmoid, for which 50\% of the asymptotic efficiency is obtained. 
}
\label{fig:hv50}
\end{center}
\end{figure}

Figure~\ref{fig:knee} 
shows the efficiency as a function of the
effective High Voltage for a typical endcap chamber. The ``knee'' is the 
$\mathrm{HV_{eff}}$ value for which 95\% of the asymptotic efficiency is reached.
The optimal working point (WP) is chosen beyond the knee, to ensure high efficiency,
and in the plateau region, to minimize the dependence on the environmental
parameters.
The WP is then defined for each individual chamber as the 
knee value plus 100 V (for barrel chambers) or 150 V (for endcap chambers).
The difference between barrel and endcap reflects the different trigger algorithms.
The endcap trigger algorithm, requiring three coincidences out of three planes,
is more sensitive to efficiency variations. 
The resulting WP values are averaged for chambers fed by the same HV supplier.

The results of the sigmoidal fit are highlighted in 
Figure~\ref{fig:hv50}, 
presenting the HV$_{50\%}$ distribution
for barrel and endcap. The different HV$_{50\%}$ average values (around 
8.9~kV for the barrel and 9.2~kV for the endcap) might be due to different
construction techniques. As shown in Figure~\ref{fig:hv50}, 
a high level of uniformity is obtained both for the barrel and for the
endcap chambers, with RMS values of the
HV$_{50\%}$ distributions of the order of 60~V and 80~V, respectively.

\section{Detector stability}
\label{section:stability}

 The variation of
the environmental pressure $\mathrm P$ and the temperature $\mathrm T$ inside the CMS cavern
was taken into account 
by rescaling~\cite{scaling} to the 
chosen reference values ($\mathrm P_0 = 965 $~mbar and
$\mathrm T_0 = 293$~K):

\begin{equation}
\nonumber
\mathrm{
HV_{eff}(P,T) = HV \cdot  \frac{P_0}{P} \cdot \frac{T}{T_0},
}
\label{eqn:corr}
\end{equation}
\noindent
where $\mathrm{HV_{eff}}$ is the resulting effective HV. 
The dominant effect is due to the pressure variation: a 1\% variation, of the order of 10~mbar,
produces a sizeable $\mathrm{HV_{eff}(P,T)}$  difference of about 100 V.
Starting from July 2011, the $\mathrm{HV_{eff}(P,T)}$  correction of 
Eq.~(\ref{eqn:corr}) was automatically implemented.

\begin{figure}[hbtp]
\includegraphics[width=.5\textwidth]{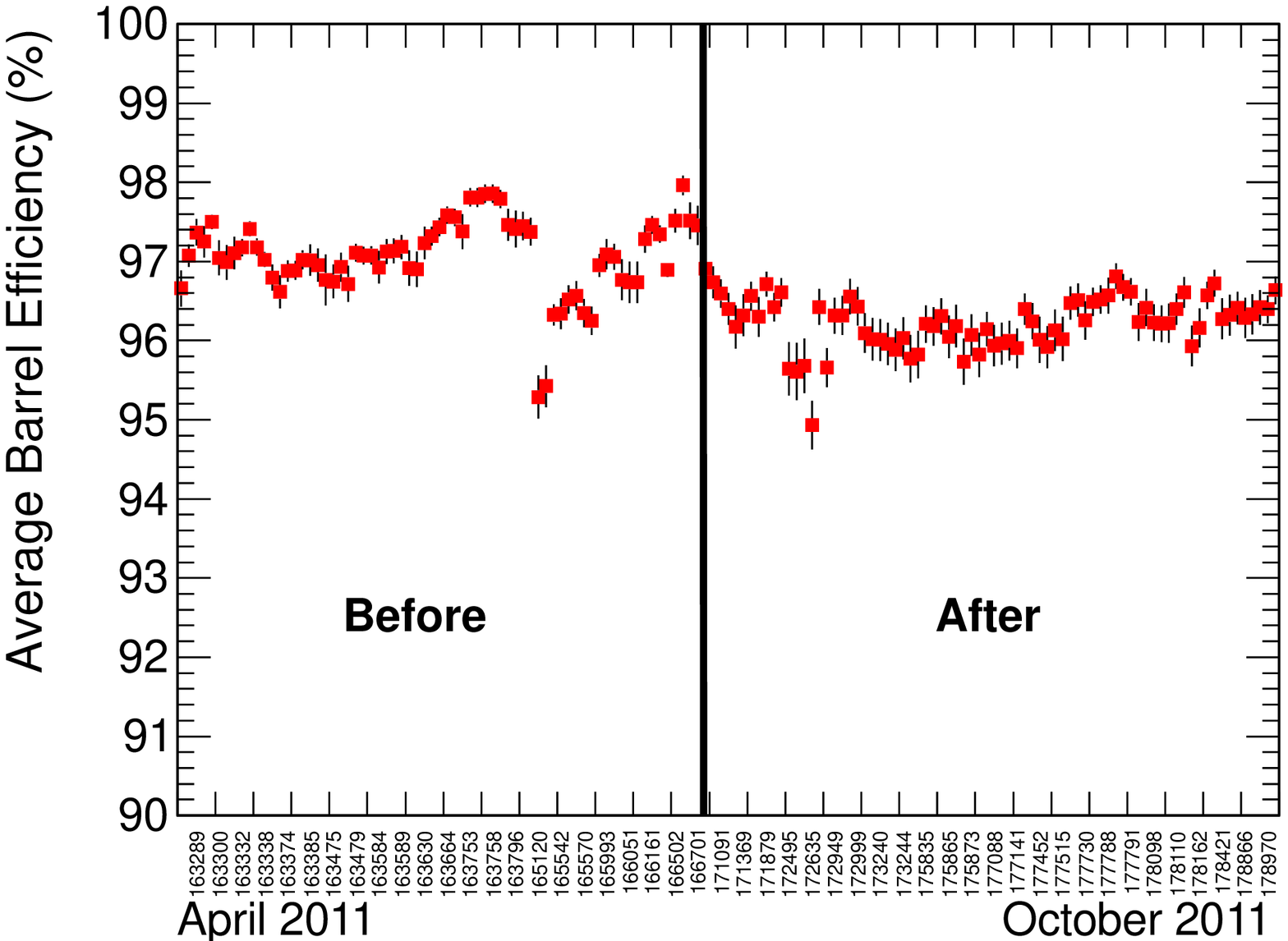} 
\includegraphics[width=.5\textwidth]{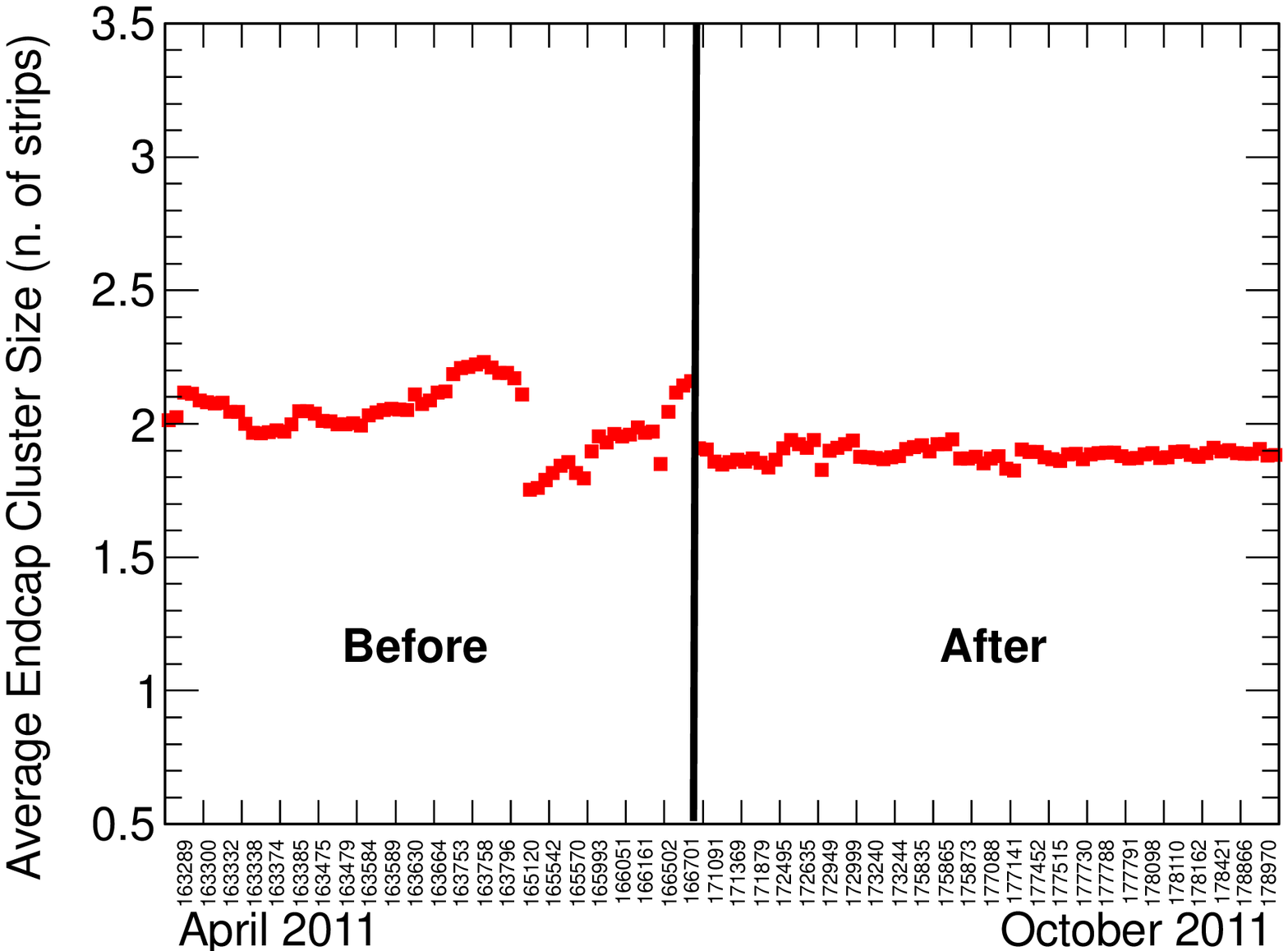} 
\caption{Left: Barrel efficiency (in \%) as a function of the run number,
{\it i.e.} as a function of time, for runs taken between April 2011 and October
2011. The two regions, before and after the automatic $\mathrm{HV_{eff}(P,T)}$  correction,
are shown in the plot.
Right: Endcap cluster size (in number of strips), as a function of the run number,
for the same run range between April and October 2011. 
}
\label{fig:eff}
\end{figure}

\begin{figure}[hbtp]
\includegraphics[width=.5\textwidth]{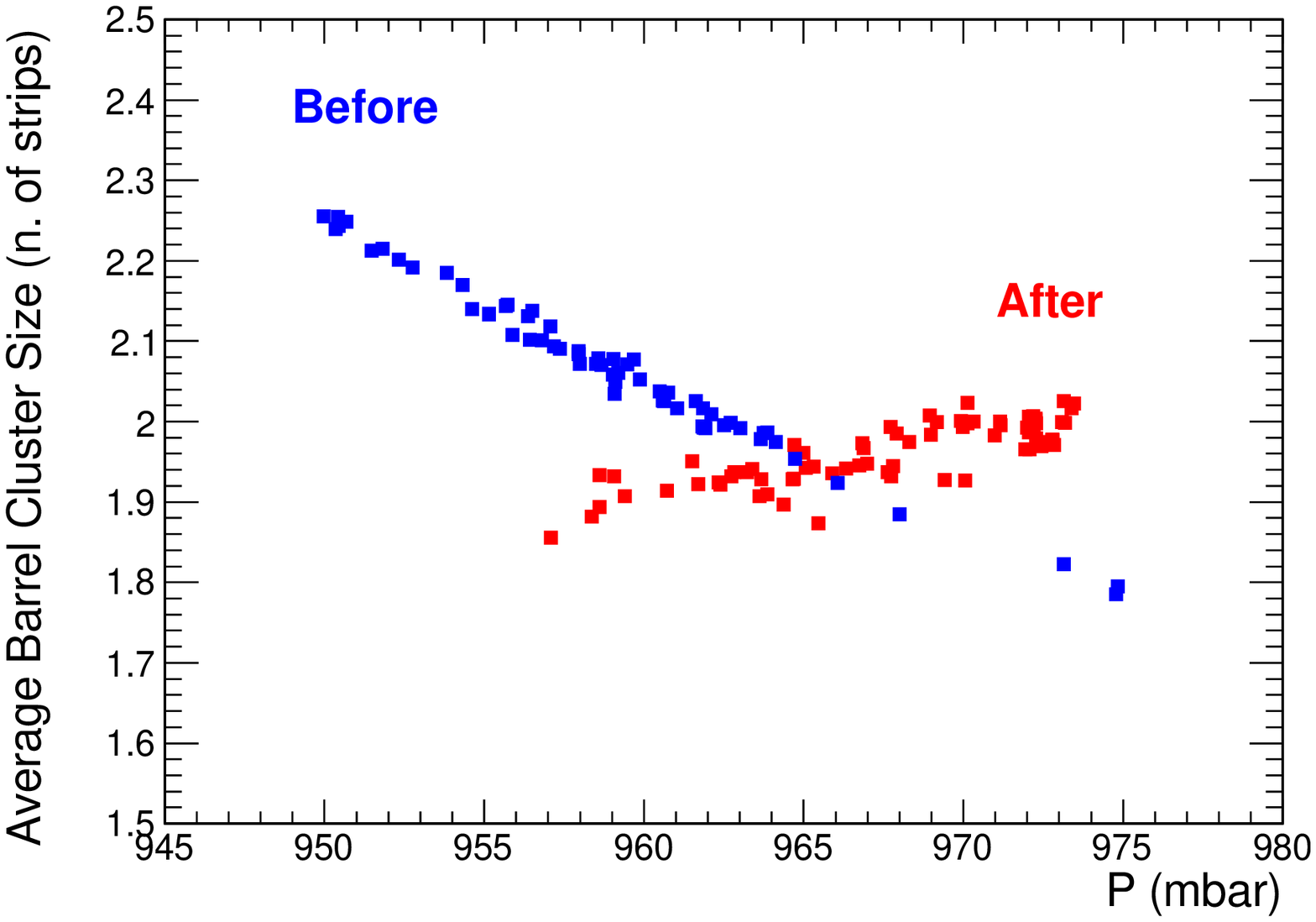} 
\includegraphics[width=.5\textwidth]{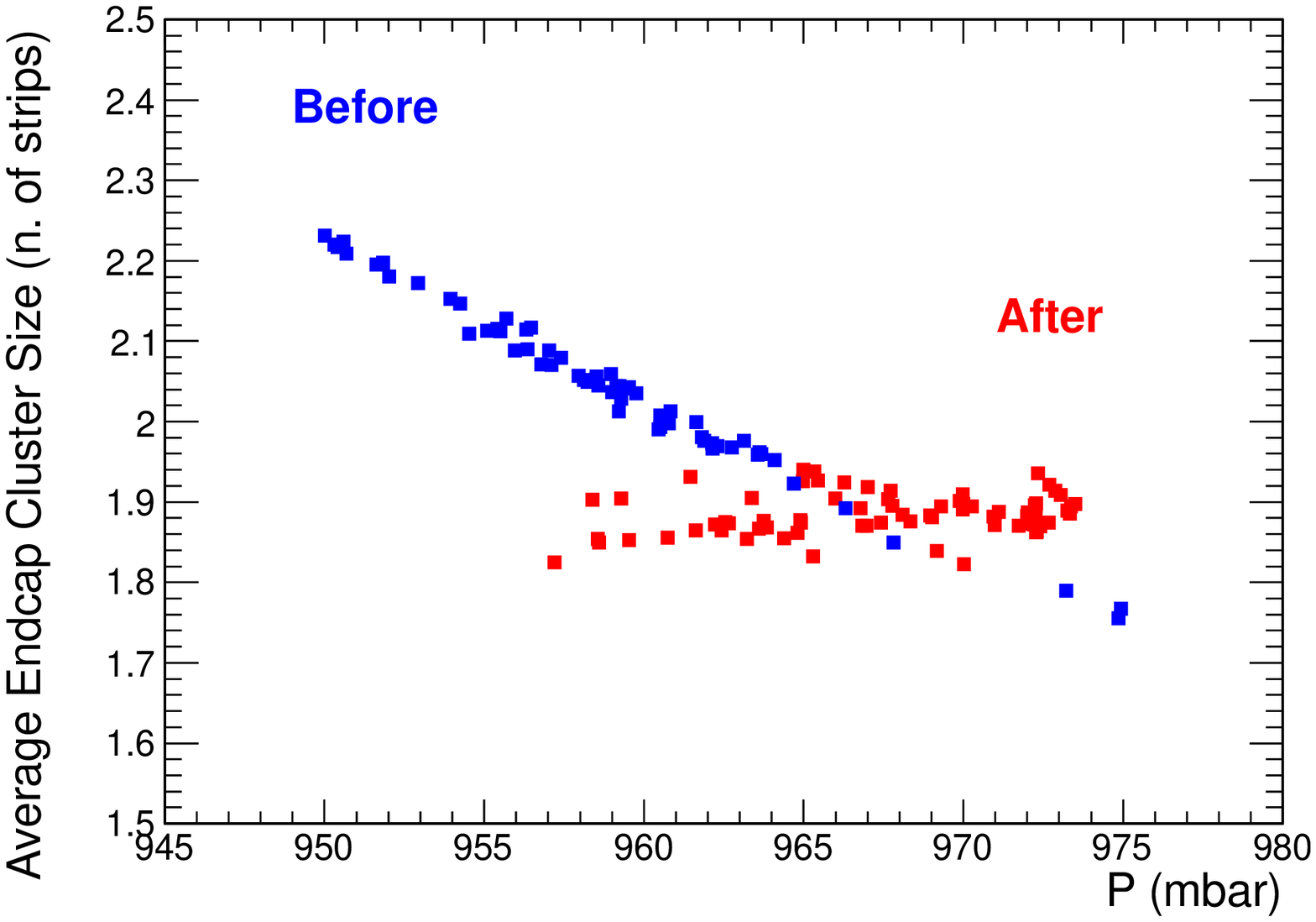} 
\caption{Average cluster size for the barrel (left) and for the endcap (right),
in number of strips, as a function of the atmospheric pressure P (in mbar),
for runs before (left, blue curve) or after (right, red curve) the
automatic $\mathrm{HV_{eff}(P,T)}$  correction. Preliminary results with
2011 data are shown.
}
\label{fig:cls}
\end{figure}

Figure~\ref{fig:eff} and Figure~\ref{fig:cls} show preliminary results
obtained with 2011 data. The efficiency and the cluster size stability
as a function of time have improved after the introduction of the
automatic $\mathrm{HV_{eff}(P,T)}$  correction of Eq.~(\ref{eqn:corr}). 
The barrel 
efficiency fluctuations (Fig~\ref{fig:eff}, left), mainly due to pressure 
variations in the CMS cavern,
are reduced from about $ \pm 1\% $ to about $ \pm 0.5\% $.
The higher average efficiency ($\sim 97\%$ compared to $\sim 96\%$) in the first
part of 2011 is due to the choice of 965 mbar as a reference value for
rescaling. In the first part of the year, (Fig~\ref{fig:eff}) the
atmospheric pressure was on average lower than 965 mbar, giving rise
to higher $\mathrm{HV_{eff}}$ values and therefore to higher efficiency 
values with respect to the second part of the year. 

An increased stability, with reduced fluctuations, is also observed in
the endcap cluster size as a function of time (Fig~\ref{fig:eff}, right).
As mentioned above for the efficiency, the same considerations apply to the average values
before and after the $\mathrm{HV_{eff}(P,T)}$  correction.

Fig~\ref{fig:cls} presents the average cluster size in the barrel (left) and in
the endcap (right) as a function of the atmospheric pressure, before and
after the automatic correction. Both in the barrel and in the endcap, a
clear anti-correlation is shown in the first part of 2011, when the correction
was not applied.
The slight positive correlation in the second part of the year might hint
at an over-correction in Eq.~(\ref{eqn:corr}), currently under study.

\section{Conclusions}

This note summarize new results highlighting the CMS RPC stability 
and uniformity over the 2011 and 2012 data taking periods.
HV scans have been performed at the beginning of 2011 and again at
the beginning of 2012. They have been extremely effective, 
allowing to select the optimal 
operating HV values for each individual RPC chamber and to obtain a high level
of uniformity both in the barrel and in the endcap.
A new method for determining the
RPC hit efficiency is used by the CMS RPC Collaboration and automatic 
$\mathrm{HV_{eff}(P,T)}$  corrections
are in place since July 2011. 
Those efforts result in increased efficiency stability and increased
cluster size stability as a function of pressure and time.

\acknowledgments

We congratulate our colleagues in the CERN accelerator departments for 
the excellent performance of the LHC machine. We thank the technical and 
administrative staff at CERN and other CMS institutes, and acknowledge 
support from BMWF and FWF (Austria); FNRS and FWO (Belgium); 
CNPq, CAPES, FAPERJ, and FAPESP (Brazil); MEYS (Bulgaria); 
CERN; CAS, MoST, and NSFC (China); COLCIENCIAS (Colombia); MSES (Croatia); 
RPF (Cyprus); MoER, SF0690030s09 and ERDF (Estonia); Academy of Finland, MEC, 
and HIP (Finland); CEA and CNRS/IN2P3 (France); BMBF, DFG, and HGF (Germany); 
GSRT (Greece); OTKA and NKTH (Hungary); DAE and DST (India); IPM (Iran); 
SFI (Ireland); INFN (Italy); NRF and WCU (Korea); LAS (Lithuania); CINVESTAV, 
CONACYT, SEP, and UASLP-FAI (Mexico); MSI (New Zealand); PAEC (Pakistan); 
MSHE and NSC (Poland); FCT (Portugal); JINR (Armenia, Belarus, Georgia, 
Ukraine, Uzbekistan); MON, RosAtom, RAS and RFBR (Russia); MSTD (Serbia); 
SEIDI and CPAN (Spain); Swiss Funding Agencies (Switzerland); NSC (Taipei); 
ThEP, IPST and NECTEC (Thailand); TUBITAK and TAEK (Turkey); NASU (Ukraine); 
STFC (United Kingdom); DOE and NSF (USA).


%
%
%
%
%

\end{document}